\documentclass[pre,twocolumn,amsmath,amssymb,showpacs,floatfix,nofootinbib]{revtex4}
\usepackage{mathptmx,times,epsfig}
\DeclareMathAlphabet\mathpzc{OT1}{pzc}{m}{it}
\let\mathcal=\mathpzc
\numberwithin{equation}{section}
\renewcommand\theequation{\arabic{section}.\arabic{equation}}

\newcommand\deriv[3][]{\frac{d^{#1}#2}{d{#3}^{#1}}}
\newcommand\partialderiv[3][]{\frac{\partial^{#1}#2}{\partial{#3}^{#1}}}
\let\trueint=\int
\let\trueiint=\iint
\let\trueiiint=\iiint
\def\int{\mathop{\textstyle\trueint}\limits}
\def\iint{\mathop{\textstyle\trueiint}\limits}
\def\iiint{\mathop{\textstyle\trueiiint}\limits}
\def\intinfty{\int\limits_{\!\!-\infty\,\,}^{\,\,\infty\!\!}\kern-0.0em}
\def\iintinfty{\mathop{\int\!\!\int}\limits_{\!\!-\infty\,\,}^{\,\,\infty\!\!}\kern-0.0em}
\def\iiintinfty{\mathop{\int\!\!\int\!\!\int}\limits_{\!\!-\infty\,\,}^{\,\,\infty\!\!}\kern-0.0em}

\def\E{{\mathbb E}}
\def\P{{\mathbb P}}
\def\F{{\cal F}}
\def\d{{\mathrm{d}}}
\def\e{{\mathrm{e}}}
\let\<=\langle
\let\>=\rangle
\let\^=\hat
\let\==\bar
\let\'=\vec
\let\~=\mathbf
\let\_=\underline
\def\sech{\mathop{\rm sech}\nolimits}
\def\Re{\mathop{\rm Re}\nolimits}
\def\Im{\mathop{\rm Im}\nolimits}

\def\Du{{\Delta u}}
\def\DL{{\Delta\@L}}
\def\dbar{{\bar d}}

\def\o#1{^{(#1)}}
\def\half{{\textstyle\frac12}}
\def\txtfrac#1#2{{\textstyle\frac{#1}{#2}}}
\catcode`\@ 11
\def\circ{\ifmmode\mathchar"220E\else$\mathchar"220E$\fi}
\catcode`\@ 13
\def\@#1{{\cal #1}}

\def\ci{{\mathrm{ci}}}
\catcode`\@ 12

\advance\textheight by 0.25cm
\advance\voffset by -0.125cm
\begin{document}
\title{Noise-induced perturbations of dispersion-managed solitons}
\author{Jinglai Li, Elaine Spiller and Gino Biondini}
\affiliation{Department of Mathematics, State University of New York at Buffalo, Buffalo, NY 14260}
\date{\small\today}

\begin{abstract}
\noindent
We study noise-induced perturbations of dispersion-managed solitons 
by developing soliton perturbation theory for the dispersion-managed 
nonlinear Schr\"odinger (DMNLS) equation, which governs the long-term
behavior of optical fiber transmission systems and certain kinds of
femtosecond lasers.
We show that the eigenmodes and generalized eigenmodes of the 
linearized DMNLS equation around traveling-wave solutions 
can be generated from the invariances of the DMNLS equations,  
we quantify the perturbation-induced parameter changes 
of the solution in terms of the eigenmodes and the 
adjoint eigenmodes, 
and we obtain evolution equations for the solution parameters. 
We then apply these results to guide importance-sampled 
Monte-Carlo simulations and reconstruct the 
probability density functions of the solution parameters
under the effect of noise. 
\end{abstract}
\maketitle

\section{Introduction}
\label{s:intro}

Dispersion management has become an essential component not only of modern
optical fiber communication systems~\cite{Agrawal,MollenauerGordon},
but also of certain femtosecond lasers~\cite{YeCundiff}.
The performance of both kinds of systems is affected by noise,
which is an essential source of system failures.  
Because these systems are designed to operate with extremely high accuracies
(typical values are one error per $10^{12}$ bits in communications 
and 1~part in~$10^{18}$ for lasers used in optical atomic clocks), 
calculating failure rates analytically is extremely difficult 
since failures result from the occurrence of unusually large 
(and therefore atypical) deviations.  
At the same time, direct Monte-Carlo computations of failure rates 
are impractical due to the exceedingly large number of 
samples that would be necessary to obtain a reliable estimate.

The effect of noise on optical transmission systems modeled by the 
nonlinear Schr\"odinger (NLS) equation has recently been studied 
\cite{OL28p105,OC256p439,PTL17p1022,PTL18p886}
using a variance reduction technique called
importance sampling (IS).
In brief, IS biases the Monte-Carlo simulations in such
a way as to artificially increase the probability of achieving
the rare events of interest, while correcting for the bias
using appropriate likelihood ratios 
(e.g., see Refs.~\cite{JLT22p1201,Srinivasan}).  
Use of IS makes it possible to efficiently estimate 
extremely small probabilities.

The key in successfully applying importance sampling lies
in biasing towards the most likely noise realizations that 
lead to system failures.
In the above-cited works, this was achieved by taking advantage 
of well-known results about the behavior of solutions of 
the NLS equation linearized around a soliton solution.
This knowledge is not available, however, in systems with
dispersion management.  The aim of this work is to address
and overcome this problem. 
We do so by employing the dispersion-managed 
NLS (DMNLS) equation which governs
the long-term dynamics of dispersion-managed optical systems
\unskip~\cite{OL23p1668,OL29p1808,OL21p327}.

The layout of this paper is as follows. 
In section~\ref{s:dmnls}
we develop a perturbation theory for dispersion-managed systems.
First we study the connection between the invariances 
of DMNLS equation and solutions of the linearized DMNLS equation.  
We then show how the equation invariances are connected to the existence
of traveling-wave solutions.
We also show that, as for the NLS equation, the linearized DMNLS around 
such traveling-wave solution can be expressed in terms of an 
ordinary differential operator
whose eigenmodes and generalized eigenmodes can also be generated 
from the same invariances. 
Finally, we use these linear modes and their adjoints 
to quantify the perturbation-induced parameter changes,
using the relation between the linear modes and the derivatives of 
the solution with respect to the invariance parameters.
In section~\ref{s:is} we use these theoretical results 
to guide importance-sampled Monte-Carlo simulations of noise-induced 
perturbations in dispersion-managed lightwave systems
and reconstruct the probability density functions of the
output solution parameters.

\section{Symmetries and perturbations of dispersion-managed systems}
\label{s:dmnls}

It is well-known that 
the propagation of coherent optical pulses in dispersion-managed systems
can be described by the following perturbed NLS equation:
\begin{equation}
i\partialderiv ut+\half\,d(t/t_a)\,\partialderiv[2]ux+g(t/t_a)|u|^2u=0\,.
\label{e:NLS+DM}
\end{equation}
Here all quantities are expressed in dimensionless units; 
$t$ is the propagation distance,
$x$ is the retarded time 
(that is, the time in a reference frame
that moves with the group velocity of the pulse), 
and $u(x,t)$ is the slowly varying envelope of the optical field,
rescaled (if necessary in communications) to take into account 
periodic loss and amplification.
The function $d(t/t_a)$ represents the local dispersion,
while $g(t/t_a)$ describes the periodic power variation due to loss 
and amplification.
(That is, the optical amplitude is proportional to 
$\sqrt{g(t/t_a)}\,u(x,t)$.)
Both $d(\,\cdot\,)$ and $g(\,\cdot\,)$ are taken to be periodic with 
unit period.
The particular choice of $d(t/t_a)$ is called a dispersion map,
and the quantity $t_a$ is called the dispersion map period.
Systems described by Eq.~\eqref{e:NLS+DM} include modern 
optical fiber communication systems~\cite{Agrawal,HasegawaKodama}
as well as certain femtosecond lasers~\cite{YeCundiff}.

\subsection{The DMNLS equation, invariances and soliton solutions}

Some of the invariances 
of the ``pure'' NLS equation (namely, Eq.~\eqref{e:NLS+DM} with $d(\cdot)=g(\cdot)=1$) are lost with dispersion management.
(More precisely, time translations, scaling and Galilean invariance,
although a generalized Galilean invariance exists.)
Moreover, Eq.~\eqref{e:NLS+DM} is a nonlinear partial differential equation
(PDE) with non-constant coefficients which contain large and rapid variations;
the asymptotic behavior of its solutions is therefore not apparent.
As shown in Ref.~\cite{OL23p1668},
an appropriate multiple-scale analysis of Eq.~\eqref{e:NLS+DM}
shows that, once the periodic compression-expansion breathing of
the pulse is properly factored out, the core pulse shape 
obeys a nonlinear, nonlocal equation of nonlinear-Schr\"odinger type
called the dispersion-managed NLS (DMNLS) equation. Without repeating
the derivation here, 
we note that the key is to split the dispersion 
$d(t/t_a)$ into the sum of two components: 
a mean value $\dbar$ and a term describing the large, zero-mean
rapid variations corresponding to the large local values of dispersion:
\begin{equation}
d(t/t_a)=\dbar+\frac1{t_a}\Delta D(t/t_a)\,.
\end{equation}
To leading order, the solution of Eq.~\eqref{e:NLS+DM} is then
\begin{gather}
\^u(\omega,t)= \^u'(\omega,t)\,e^{-iC(\zeta)\omega^2/2},
\label{e:solndecomposition}
\\
\noalign{where $\zeta=t/t_a\,$ and}
C(\zeta)=C_0+\int_0^\zeta \Delta D(\zeta')\,d\zeta'\,,
\label{e:Czeta}
\end{gather}
with $C_0$ an arbitrary integration constant.
Above and hereafter, 
$\^f(\omega)=\F_\omega[f(x)]= \int \e^{-i\omega x}f(x)\,\d x$
is the Fourier transform of $f(x)$.
The exponential factor in Eq.~\eqref{e:solndecomposition} 
accounts for the rapid breathing of the pulse, while the
slowly varying envelope $\^u(\omega,t)$ satisfies the DMNLS equation, 
which in the physical and Fourier domains, is respectively 
(omitting primes for simplicity)
\begin{subequations}
\label{e:DMNLS}
\begin{multline}
i\partialderiv ut+\half\dbar\partialderiv[2]ux
\\ \kern4em{ }
  +\iint u^{}_{(x+x')}u^{}_{(x+x'')}u^*_{(x+x'+x'')} 
     R^{}_{(x',x'')}\,\,dx'dx'' = 0\,
\label{e:DMNLSx}
\end{multline}
\begin{multline}
i\partialderiv{\^u}t-\half\dbar\omega^2\^u
\\ \kern2em{ }
  +\iint 
    \^u^{}_{(\omega+\omega')}\^u^{}_{(\omega+\omega'')}\^u^*_{(\omega+\omega'+\omega'')}
    r_{(\omega'\omega'')}\,\, d\omega'd\omega'' = 0\,,
\label{e:DMNLSf}
\end{multline}
\end{subequations}
where the asterisk denotes complex conjugation and where
for brevity we introduced the shorthand notations 
$u_{(x)}=u(x,t)$, $\^u_{(\omega)}=\^u(\omega,t)\,$ etc.
Throughout this work, integrals are complete unless limits 
are explicitly stated.
The integration kernels $r(y)$ and $R(x',x'')$ in Eqs~\eqref{e:DMNLS}
quantify the average nonlinearity over a dispersion map
mitigated by the dispersion management, and are given respectively by 
\begin{subequations}
\begin{gather}
r(y)= \frac1{(2\pi)^2 }\int_0^{1} g(\zeta)e^{iC(\zeta)y}\,d\zeta\,,
\label{e:rdef}
\\
R(x',x'')=\frac1{2\pi}
  \iint e^{-i\omega'x'-i\omega''x''}r(\omega'\omega'')\,d\omega'd\omega''\,.
\label{e:Rdef}
\end{gather}
\end{subequations}
Note that both focusing and defocusing cases can be obtained 
in Eqs.~\eqref{e:DMNLS}
depending on the sign of the average dispersion~$\dbar$.

It is crucial to realize that the DMNLS equation
and its solutions depend implicitly on a parameter, called the 
dimensionless \textit{reduced map strength}, 
which quantifies the size of the zero-mean dispersion fluctuations.
The map strength $s$ can be defined for any dispersion map as
\begin{equation}
s= \txtfrac14\|\Delta D\|_1=
  \txtfrac14\int_0^1|\Delta D(\zeta)|\,d\zeta\,.
\label{e:sdef}
\end{equation}
One can then formally obtain the dependence of the 
kernels $r(y)$ and $R(x',x'')$ on the map strength
by writing $\Delta D(\zeta)$ and $C(\zeta)$ in terms of
normalized functions which only depend on the shape of 
the zero-mean dispersion variations. 
Namely, one writes
%
$\Delta D(\zeta)= 4s\,\Delta D_1(\zeta)$ and 
$C(\tau)=4s\,C_1(\tau)$\,,
where $C_1(\zeta)$ is given by Eq.~\eqref{e:Czeta} with 
$\Delta D(\zeta)$ replaced by $\Delta D_1(\zeta)$.
In this way, one can conveniently study cases with different map strengths 
entirely within the framework of the DMNLS equation,
without needing to refer to Eq.~\eqref{e:NLS+DM}. 
Of course, in the limit $s\to0$ one obtains $r(y)\to1/(2\pi)^2$,
and $R(x',x'')\to\delta(x')\delta(x'')$.
That is, as $s\to0$ the DMNLS equation~\eqref{e:DMNLS} 
reduces to the pure NLS equation. 

The DMNLS equation~\eqref{e:DMNLS} is a reduced model that retains 
the essential features of dispersion-managed systems while bypassing the 
complicated dynamics that take place within each dispersion map.
As such, it has proved to be a useful model to investigate 
the long-time behavior of dispersion-managed systems
\cite{OL23p1668,JOSAB18p577,OL26p459,JOSAB19p2876,OL29p1808,OL26p1761,%
Capobianco,OL21p327,PLA260p68,CHA10p539,OL25p1144,OL26p1535,PRE62p4283,Qurashi,Tonello:OC1,Tonello:OC2}.

In the case where loss and gain are perfectly balanced
(e.g., with distributed Raman amplification in communication
systems~\cite{PRL96p023902}) 
it is $g(\,\cdot\,)=1$, and both kernels can then be made real
by proper choice of the integration constants. 
Here we assume that this has been done. 
Then, in the special but physically important case of a piecewise constant, 
two-step dispersion map, the kernels assume a particularly simple form
\unskip~\cite{OL23p1668}:
\begin{equation}
r(y)= \frac1{(2\pi)^2}\frac{\sin sy}{sy},
\qquad
R(x',x'')= \frac1{2\pi|s|}\,\ci({x'x''}/s)\,,
\end{equation}
where $\ci(x)=\int\nolimits^{\infty}_1 \cos(xy)/y~dy\,$.
Note that in this case both kernels are independent of 
the particular shape of the zero-mean dispersion variations.
The same kernels, apart from a factor 2, also arise for the DMNLS equation 
as a model of certain femtosecond lasers~\cite{OL29p1808}.

Stationary solutions of the DMNLS equation are obtained by 
looking for solutions of the form:
\begin{equation}
u_{\mathrm{st}}(x,t;s)= e^{i\lambda^2t/2}f(x;s)\,.
\label{e:stationary}
\end{equation}
The Fourier transform $\^f(\omega)$ of $f(x)$ then 
solves the nonlinear integral equation~\cite{OL23p1668}\,
\begin{multline}
\big(\lambda^2+\dbar\omega^2\big)\,\^f_{(\omega)} =
\\ 
  2\iint 
    \^f{}_{(\omega+\omega')}\^f{}_{(\omega+\omega'')}\^f^*_{(\omega+\omega'+\omega'')}
    r_{(\omega'\omega'')}\,\, d\omega'd\omega''\,,
\label{e:DMNLSintleqn}
\end{multline}
which can be efficiently solved numerically
(cf.\ Appendix \ref{s:A2}).
Note that, like in the NLS equation, the ``soliton eigenvalue''~$\lambda$ 
is also the peak amplitude of the pulse,
independently of map strength.

Remarkably, some of the invariances of the standard NLS equation that
are destroyed by dispersion management are recovered by the DMNLS 
equation.
In particular, the following NLS invariances also hold
for the DMNLS equation without modification: 
phase invariance,
position invariance, 
time invariance
and Galilean invariance,
which are respectively 
\begin{subequations}
\label{e:invar}
\begin{gather}
u\to e^{i\phi_o}u\,,\\
x\to x-x_o\,,\\
t\to t-t_o\,,\\
x\to x-\Omega t,\qquad  u\to e^{i\Omega x - i\Omega^2 t/2} u\,. 
\end{gather}
On the other hand, unlike the NLS equation, the DMNLS equation 
is not scale invariant.
Rather, the DMNLS equation admits a generalized scaling invariance
which also involves the map strength:
\begin{gather}
x\to x/A,\qquad t\to t/A^2,\qquad s\to s/A^2,\qquad u\to Au.
\label{e:scale}
\end{gather}
\end{subequations}
Starting from Eq.~\eqref{e:stationary},
a four-parameter family of traveling-wave solutions can 
then be generated by using the invariances \eqref{e:invar}: 
\begin{subequations}
\label{e:trvsln}
\begin{equation}
u(x,t)= A\,\e^{i\,[\Omega x+\half(A^2-\Omega^2)t + \phi_o]}
  f(A(x-x_o-\Omega t);A^2s)\,,
\label{e:trvsln1}
\end{equation}
or, equivalently,
\begin{equation}
u(x,t)=A\,\e^{i(\Omega x+\Phi)}f(A(x-X);A^2s)\,,
\label{e:trvsln2}
\end{equation}
\end{subequations}
where 
\begin{equation}
X(t)=x_o+\Omega t\,,\qquad
\Phi(t)=(A^2-\Omega^2)t/2+\phi_o\,
\end{equation}
are respectively the mean position and an overall phase. 
Note that for the stationary solution \eqref{e:stationary}, 
time translations can be expressed as a composition of 
phase transformations and position translations. 
Hence, even though five invariances exist, 
there are only four independent solution parameters
for traveling-wave solutions. 
When the kernel $r(y)$ is real, $f(x)$ can be taken to be real and even.
Solutions~\eqref{e:trvsln} are usually referred as 
dispersion-managed solitons (DMS).

\subsection{Linear modes of the DMNLS equation}

We now consider the stability of solutions under perturbations.
If $u(x,t)$ is any solution of the DMNLS equation and 
$u(x,t)+\epsilon v(x,t)$ is also a solution, 
$v(x,t)$ solves the linearized DMNLS equation around $u(x,t)$;
namely, $L[v,u]=0\,$,
where \cite{PRE62p4283,CHA10p539}
\begin{multline}
L[v,u]=\partialderiv vt - \txtfrac i2 \dbar\,\partialderiv[2]vx
 \\{} - 2i\iint u_{(x+x'')}u^*_{(x+x'+x'')}v^{}_{(x+x')}R_{(x',x'')}\,\,dx'dx''
\\{ }
  - i\iint u^{}_{(x+x')}u^{}_{(x+x'')}v^*_{(x+x'+x'')}R_{(x',x'')}\,\,dx'dx''\,
\label{e:LDMNLS}
\end{multline}
is the linearized DMNLS operator. 
Since the DMNLS equation is not integrable
\cite{JOSAB18p577,PLA260p68},
its linear modes cannot be derived from the inverse scattering method 
as for the NLS equation~\cite{Kaup:1991}.
The linear modes, however, can be generated using 
the invariances~\eqref{e:invar}.
Suppose that $u(x,t)$ solves some PDE,
and consider a generic infinitesimal transformation 
$u(x,t)\to u_\epsilon(x,t)$, with
\begin{subequations}
\begin{gather}
u_\epsilon(x,t)=u(x,t)+\epsilon v(x,t) + O(\epsilon ^2)\,,
\label{e:utransf}
\\
\noalign{and}
v(x,t)=\partialderiv{u_\epsilon(x,t)}\epsilon\bigg |_{\epsilon=0} \,.
\label{e:mode}
\end{gather}
\end{subequations}
If the PDE is invariant under the above transformation, 
one verifies that $v(x,t)$ is in the nullspace of $L[\,\cdot\,,u]$;
namely, $v(x,t)$ is a solution of the linearized PDE about 
the given solution.
When applied to the DMNLS equation, this construction
yields four solutions of the linearized DMNLS equation around 
any solution~$u(x,t)$.
More precisely, these solutions of the linearized DMNLS equation, 
which are associated with the phase, distance translation,
Galilean invariance and scale invariance, are respectively
\begin{subequations}
\label{e:Lsolutions}
\begin{gather}
v_1 = iu\,,\qquad
v_2 = -\partialderiv ux\,,
\label{e:linmodes}
\\
v_3 = ix\,u-t\partialderiv ux\,,\qquad
v_4 = u+x\partialderiv ux+2t\partialderiv ut+2s\partialderiv us\,.
\label{e:genlinmodes}
\end{gather}
\end{subequations}
Note that $v_3$ and $v_4$ are not bounded in time.
Using the fact that $L[v]=w$ implies $L[tv]= tw + v$,
however,
one can convert $v_3$ and $v_4$ into bounded elements of
the generalized nullspace of~$L$.
Note also that a further solution of the linearized DMNLS equation,
namely $v_5=-u_t$, 
can be generated from invariance with respect to time translations. 
This fifth solution of the linearized equation, however, 
is \textit{not} linearly independent from the other four
if $u(x,t)$ is the traveling-wave solution~\eqref{e:trvsln}, 
since then
\begin{equation}
v_5=\half(A^2+\Omega^2)\,v_1+\Omega\, v_2\,.
\label{e:vHlincomb}
\end{equation}

\begin{figure}[t!]
\centerline{\includegraphics[width=0.45\textwidth]{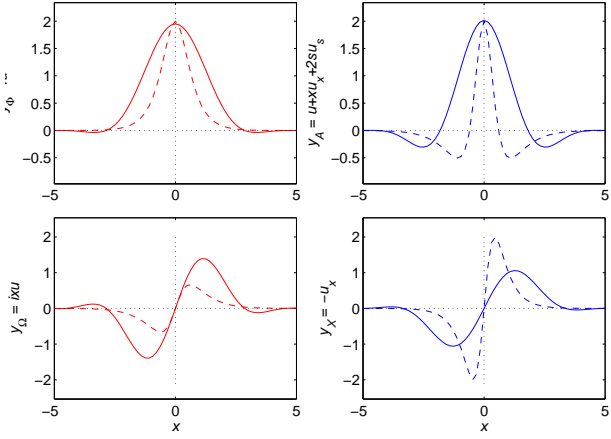}}
\caption{The shape of the neutral eigenmodes and generalized eigenmodes 
of the DMNLS equation for $s=2$ (solid lines),
compared to the corresponding modes for the NLS equation
($s=0$, dashed lines),
with $\dbar=1$ and $A=2$ in both cases.
Blue lines (top right and bottom right plots) 
show components in phase with the pulse;
red lines (top left and bottom left plots)
show components 90$^\circ$ out of phase.}
\label{f:DMNLSmodes}
\kern-\medskipamount
\end{figure}

For traveling-wave solutions, it is possible to express the 
linearized DMNLS equation in terms of an ordinary differential operator
by performing a change of coordinates to the comoving frame $(\xi,t')$, 
with $\xi=x-X(t)\,$ and $t'=t\,$,
and writing $u(x,t)$ and $v(x,t)$ respectively as
\begin{equation}
u(x,t)=e^{i\Theta}U(\xi)\,,\qquad
v(x,t)=e^{i\,\Theta}y(\xi,t')\,,
\label{e:ordinary}
\end{equation}
where $\Theta(x,t)=\Omega x+\Phi(t)$ is the local phase
and $U(\xi)=Af(A\xi)$ the pulse envelope. 
Substituting Eq.~\eqref{e:ordinary} into Eq.~\eqref{e:LDMNLS} yields
\begin{equation}
e^{-i\Theta}L[v;u]=\partialderiv{y}{t'}-\Lambda[y;U]\,,\label{e:LLambda}
\end{equation}
where 
\begin{multline}
\Lambda[y,U]=
   \frac i2\dbar\,\partialderiv[2]{y}\xi -\frac i2A^2y\\
   + 2i\iint U_{(\xi+\xi'')}U^*_{(\xi+\xi'+\xi'')}y^{}_{(\xi+\xi')}R_{(\xi',\xi'')}\,\,d\xi'd\xi''\\
  +i\iint U_{(\xi+\xi')}U_{(\xi+\xi'')}y^*_{(\xi+\xi'+\xi'')}R_{(\xi',\xi'')}\,\,d\xi'd\xi''\,.
\label{e:odeoperator}
\end{multline}
%
Following standard terminology, 
we will call $y$ a neutral eigenmode if $\Lambda[y]=0$ 
and a generalized eigenmode if $\Lambda[y]$ is a neutral eigenmode.
Using Eqs.~\eqref{e:ordinary} and \eqref{e:LLambda},
one can associate each solution of the linearized DMNLS 
in Eqs.~\eqref{e:Lsolutions} 
to a neutral eigenmode or generalized eigenmode of $\Lambda$.
After rearranging terms, one can obtain the following 
set of modes and generalized modes:
\begin{subequations}
\label{e:Lambdamodes}
\begin{gather}
y_\Phi= iU\,,\qquad
y_X= -\partialderiv U\xi\,,
\label{e:linmodes2}
\\
y_\Omega= i\xi\,U\,,\quad
y_A = \frac1A\bigg(U+\xi \partialderiv U\xi+2s\partialderiv Us\bigg)\,
\label{e:genlinmodes2}
\end{gather}
\end{subequations}
(where the subscript associates each mode to the solution parameters changed by
the transformation), which satisfy the following relations:
\begin{subequations}
\label{e:Lambdaeqns}
\begin{gather}
\Lambda[y_\Phi]=0\,,\qquad 
\Lambda[y_X]=0\,,
\\
\Lambda[y_\Omega]= y_X\,,\qquad
\Lambda[y_A]= A\,y_\Phi\,.
\end{gather}
\end{subequations}
Note the explicit dependence on $s$ of the amplitude mode $y_A$,
as well as, of course, of the corresponding solution of the 
linearized DMNLS equation, $v_4$.
Note also that, as for the NLS equation, some freedom exists 
in the definition of the generalized modes,
as well as in the normalization of all modes.
Importantly, for $Q=A,X,\Phi$ it is
\begin{subequations}
\label{e:modes&derivs}
\begin{gather}
\partialderiv u{Q}=e^{i\Theta}y_Q\,,
\\[-0.4ex]
\noalign{while}
\partialderiv u\Omega= e^{i\Theta}\,(y_\Omega+X\,y_\Phi)\,.
\label{e:dudOmega}
\end{gather}
\end{subequations}
Of course, different parametrization of the traveling-wave solution
will also result in different combinations of modes.
The shape of the modes in Eqs.~\eqref{e:Lambdamodes} 
is shown in Fig.~\ref{f:DMNLSmodes} 
for $s=0$ (NLS) and $s=2$.

\begin{figure}[t!]
\includegraphics[width=0.405\textwidth]{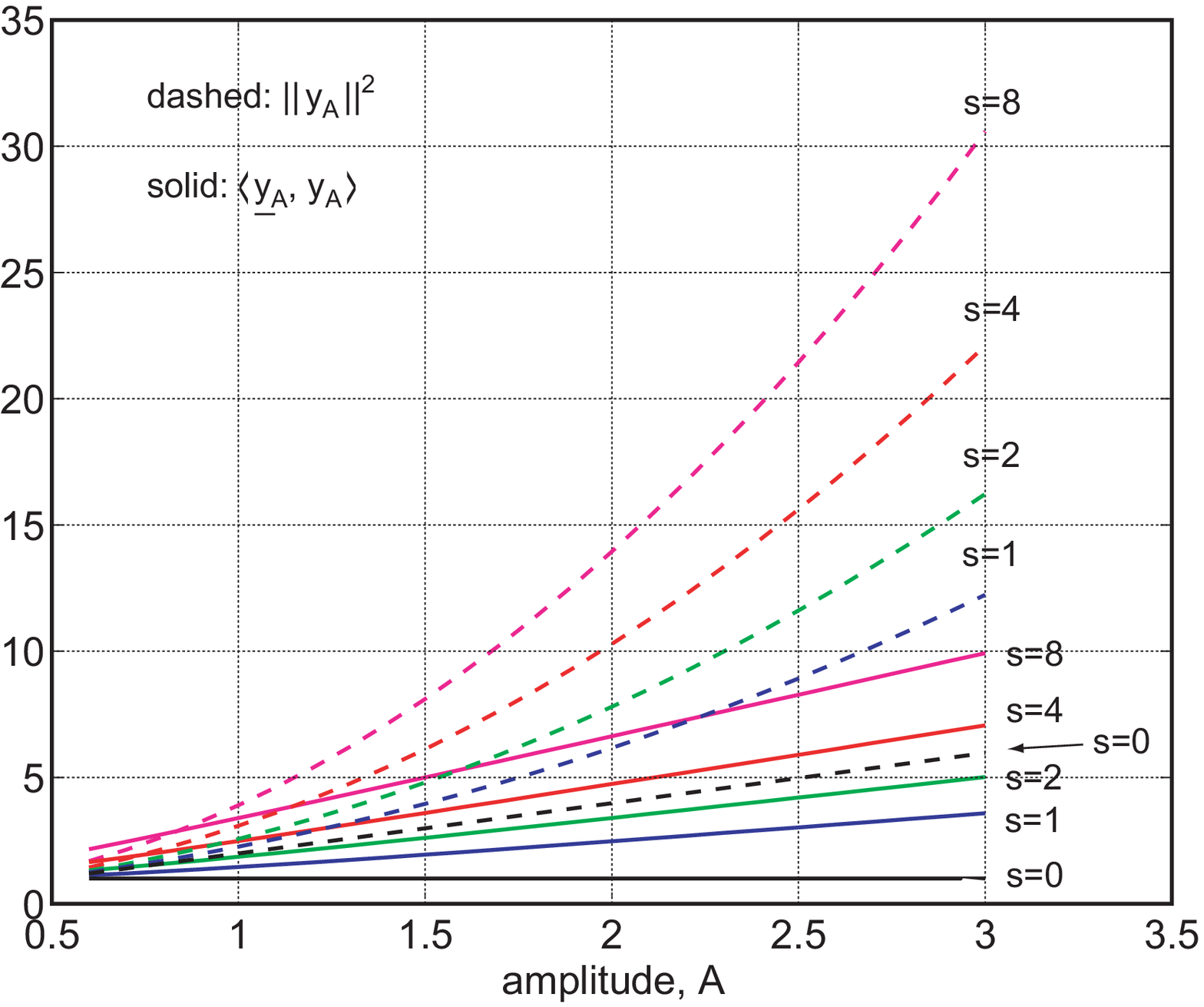}
\vglue\bigskipamount
\includegraphics[width=0.405\textwidth]{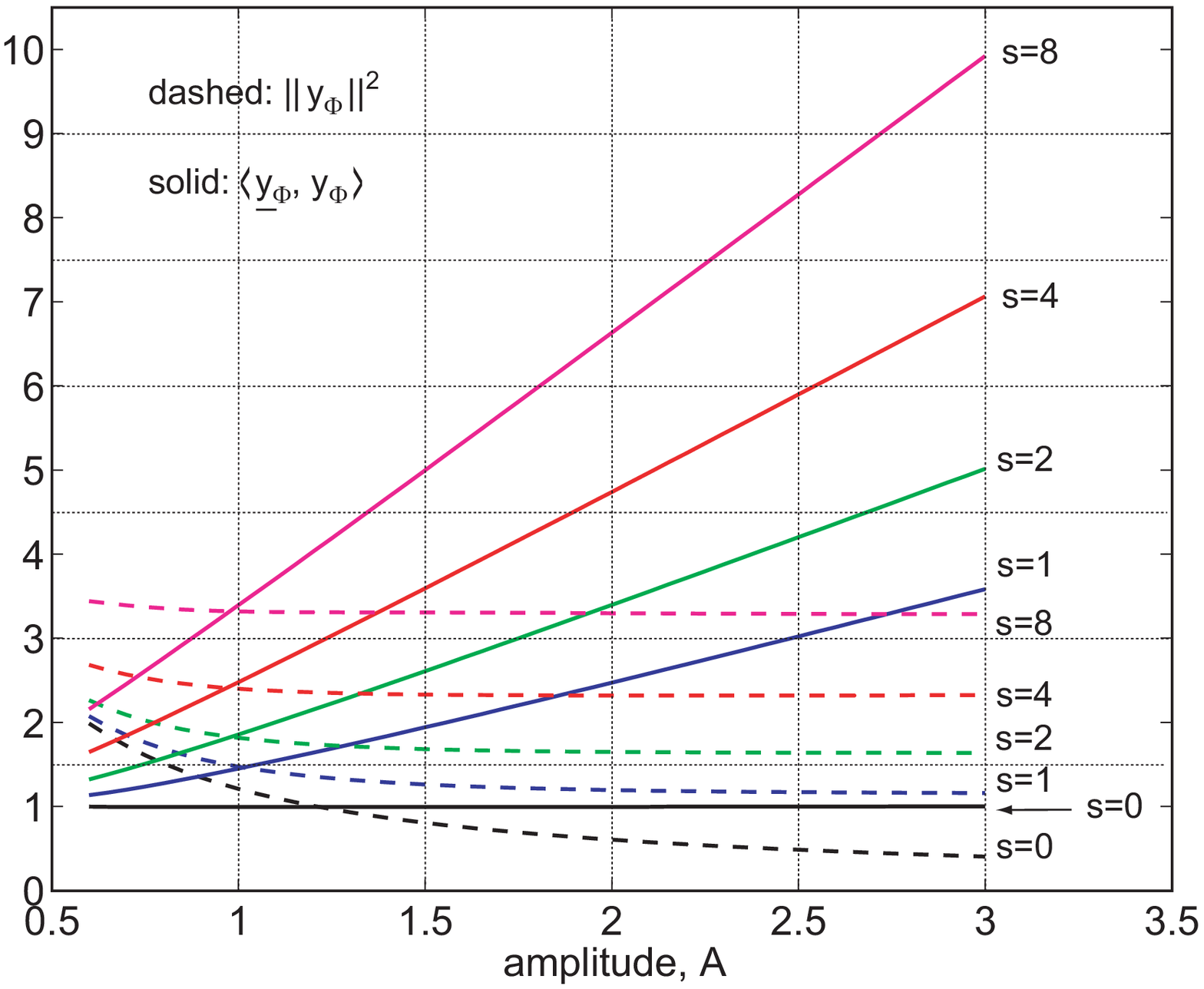}
\caption{Top: the $L_2$-norm of  of the amplitude mode $\_y_A$ 
and the inner product $\<\_y_A,y_A\>$ as a function of the 
amplitude parameter~$A$ for different values of map strength. 
Bottom: the $L_2$-norm of $\_y_\Phi$ and the inner product 
$\<\_y_\Phi,y_\Phi\>$ as a function of the amplitude parameter~$A$.}
\label{f:modenorms}
\kern-\medskipamount
\end{figure}

In order to quantify the effect of perturbations, it is necessary
to also employ the adjoint modes.
Introducing the inner product as 
\begin{equation}
\<y,w\>
= \Re\int y^*(x)w(x)\,dx
= \int(y_{\Re}w_{\Re}+y_{\Im}w_{\Im})\,\,dx\,,
\end{equation}
%
the adjoint of $\Lambda[y,U]$ is found to be
\begin{multline}
\Lambda^\dagger[y,U]=
  - \frac i2\dbar\,\partialderiv[2]{y}\xi +\frac i2A^2y\\
   - 2i\iint U_{(\xi+\xi'')}U^*_{(\xi+\xi'+\xi'')}y^{}_{(\xi+\xi')}R_{(\xi',\xi'')}\,\,d\xi'd\xi''\\
  +i\iint U^{}_{(\xi+\xi')}U^{}_{(\xi+\xi'')}y^*_{(\xi+\xi'+\xi'')}R_{(\xi',\xi'')}\,\,d\xi'd\xi''\,. 
\label{e:adjodeop}
\end{multline}
Note that
$\Lambda^\dagger[y,U]=-i\Lambda[iy,U]\,$.
Using this property, one can immediately obtain 
the adjoint modes of Eqs.~\eqref{e:Lambdamodes}, 
i.e., the eigenmodes of~$\Lambda^\dagger$:
\begin{subequations}
\label{e:adjmodes}
\begin{gather}
\_y_\Phi= \frac iA\bigg(U+\xi \partialderiv U\xi+2s\partialderiv Us\bigg)\,,\quad
\_y_X= \frac1A\,\xi\,U\,,
\\
\_y_\Omega= -\frac iA\,\partialderiv U\xi\,.\quad
\_y_A= U,
\end{gather}
\end{subequations}
satisfying relations analogous to Eqs.~\eqref{e:Lambdaeqns}.
(Note, however, that $\_y_Q\ne \pm i y_Q$, and 
the adjoint of a neutral eigenmode is a generalized eigenmode, 
and viceversa.)\,\ 
Due to the real and even nature of $f(x)\,$, 
the modes~\eqref{e:Lambdamodes} and their adjoints~\eqref{e:adjmodes}
form a bi-orthogonal set which provides
a basis of the generalized nullspace of $\Lambda$.
Explicitly,
\begin{equation}
 \<\_y_Q\,,y_{Q'}\>= \<\_y_Q\,,y_Q\>\,\delta_{QQ\,'}\,,
\end{equation}
where 
\begin{subequations}
\begin{gather}
\<\_y_A,y_A\>= \<\_y_\Phi,y_\Phi\>=
  \left(\half+2s\partialderiv{ }s\right)E/A\,,\kern-1em{ }\\
\<\_y_X,y_X\>=\<\_y_\Omega,y_\Omega\>= \half\, E/A\,,\kern1em
\end{gather}
\end{subequations}
and where $E= \|u\|^2= \int |u(x,t)|^2\,dx$ is the pulse energy. 
Note that, unlike the case of the NLS equation,
these adjoint modes are not normalized, 
and in general the norms $\smash{\|y_Q\|^2=\<\_y_Q, y_Q\>}$ 
depend on both $s$ and $\lambda$, 
as shown in Fig~\ref{f:modenorms}.
Of course the orthonormality could be achieved by properly redefining 
the adjoint modes, but the present choice is convenient for our purposes.
For the NLS equation it is simply $s=0$ and $E=2A$,
so in that case the modes are indeed bi-orthonormal.
As we show next,
these linear modes and adjoint modes can  be used to quantify 
perturbation-induced solution parameter changes.

\subsection{Perturbation-induced parameter changes}
\label{s:change}

We now consider perturbations which manifest
as an additional term in the NLS equation~\eqref{e:NLS+DM}:
\begin{equation}
i\partialderiv ut+\half d(t/t_a)\partialderiv[2]ux\\
  +g(t/t_a)|u|^2u = i\epsilon h(x,t)\,,
\label{e:PNLS+DM}
\end{equation}
with $0<\epsilon\ll1$.
Using the same multiple scale analysis as for Eq.~\eqref{e:NLS+DM},
from Eq.~\eqref{e:PNLS+DM} one obtains a perturbed DMNLS equation
in which an inhomogeneous term is added to the right-hand side (RHS):
\begin{multline}
i\partialderiv ut+\half\dbar\partialderiv[2]ux\\
  +\iint u^{}_{(x+x')}u^{}_{(x+x'')}u^*_{(x+x'+x'')} 
     R^{}_{(x',x'')}\,\,dx'dx'' = i\epsilon h(x,t)\,.
\label{e:PDMNLS}
\end{multline}
This formulation is of course very general, 
and it includes most physically interesting situations such as 
damping, amplification, third-order dispersion, shock and Raman effects 
(e.g., see Refs.~\cite{Agrawal,HasegawaKodama}).

Suppose $u_\epsilon=u+\epsilon v$ solves Eq.~\eqref{e:PDMNLS}, 
where $u$ is a traveling-wave solution of the unperturbed DMNLS 
equation given by Eq.~\eqref{e:trvsln}\, and $\epsilon v$ 
is the perturbation-induced solution change.
Then the perturbation $v(x,t)$ satisfies the perturbed linearized
DMNLS equation, 
\begin{equation}
L[v]=h\,.
\label{e:PLDMNLS}
\end{equation}
Of course, in order for the perturbation expansion to remain 
well-ordered, the solutions of Eq.~\eqref{e:PLDMNLS} must remain bounded.
If, however, the right-hand-side of Eq.~\eqref{e:PLDMNLS} 
has a non-zero component along the nullspace of~$L$, 
secular terms will arise.
As usual, such terms are removed by requiring that 
the parameters of the unperturbed solution become 
slowly-varying functions of time. 
Namely, introducing
the fast and slow time scales $t_1=t$ and $t_2=\epsilon t$,
one obtains
\begin{equation}
\label{e:uexp}
\partialderiv{u}t=\partialderiv{u}{t_1}+ \epsilon
\sum_Q \partialderiv{u}{Q} \partialderiv{Q}{t_2}\,,
\end{equation}
where now $Q=A,\Omega,X,\Phi$.
The perturbed linearized DMNLS equation~\eqref{e:PLDMNLS} therefore becomes
\begin{equation}
L_1[v]+ \sum_Q \partialderiv{u}{Q}\partialderiv{Q}{t_2} = h\,,
\label{e:pdq}
\end{equation}
where $L_1[v]$ is given by Eq.~\eqref{e:LDMNLS} with $\partial/\partial t$
replaced by $\partial/\partial t_1$.
Recalling Eqs.~\eqref{e:LLambda} and~\eqref{e:modes&derivs},
one can rewrite Eq.~\eqref{e:pdq} as
\begin{subequations}
\begin{equation}
\partialderiv {w}{t_1} - \Lambda_1[w]
  + Xy_\Phi\,\partialderiv\Omega{t_2}
  + \sum_Q y_Q\partialderiv{Q}{t_2} = \e^{-i\Theta}h\,,
\label{e:wexpmodes}
\end{equation}
\end{subequations}
where $w(x,t)=e^{-i\Theta(x,t)}v(x,t)$
and where $\Lambda_1[w]$ is given by Eq.~\eqref{e:odeoperator}
with $t'=t_1$ and $\xi=x-X(t_1)$.
[The extra term proportional to $y_\Phi$ comes from Eq.~\eqref{e:dudOmega}.]\,\ 
We now decompose the right-hand side of Eq.~\eqref{e:wexpmodes} as
\begin{subequations}
\begin{gather}
h =
  \sum_Q
    \frac{\<e^{i\Theta}\_y_Q,h\>}{\<\_y_Q,y_Q\>}e^{i\Theta}\,y_Q
      + h_{\mathrm{res}}\,,
\label{e:pexp}
\\
\noalign{so that}
\<e^{i\Theta}\_y_Q,h_{\mathrm{res}}\>=0
\end{gather} 
\end{subequations} 
for $Q=A,\Omega,X,\Phi$.
%
%
The solvability condition 
(namely the requirement that the resulting PDE for $w$ contain
no secular terms) 
then provides evolution equations for the solution parameters:
\begin{subequations}
\label{e:paramderivs}
\begin{gather}
\deriv{A}t= \epsilon\,\frac{\<e^{i\Theta}\_y_A,\,h\>}{\<\_y_A,y_A\>}\,,
\\
\deriv{\Omega}t= \epsilon\,\frac{\<\e^{i\Theta}\_y_\Omega,\,h\>}{\<\_y_\Omega,y_\Omega\>}\,,
\\
\deriv{X}t= \Omega+\epsilon\,\frac{\<e^{i\Theta}\_y_X,\,h\>}{\<\_y_X,y_X\>}\,,
\\
\deriv{\Phi}t= \half(A^2-\Omega^2)+\epsilon\,\frac{\<e^{i\Theta}\_y_\Phi,\,h\>}{\<\_y_\Phi,y_\Phi\>}
 - \epsilon\,X\,\frac{\<\e^{i\Theta}\_y_\Omega,\,h\>}{\<\_y_\Omega,y_\Omega\>}\,.
\end{gather}
\end{subequations}
In the special case $h(x,t)= \Delta u(x)\,\delta(t-t_o)$,
Eq.~\eqref{e:PDMNLS} describes changes in the initial condition.
[Integrating from $t=t_o-\Delta t$
to $t=t_o+\Delta t$ and letting $\Delta t\to0$ one has
$u(x,t_o^+)= u(x,t_o^-)+\epsilon\,\Delta u(x)$.]\,\ 
In this case, Eqs.~\eqref{e:paramderivs} yield
the parameter changes as 
$Q(t_o^+)= Q(t_o^-)+\epsilon\,\Delta Q\,$, where
\begin{subequations}
\label{e:intlchng}
\begin{gather}
\Delta Q=\frac{\<e^{i\Theta}\_y_Q,\,\Du\>}{\<\_y_Q,y_Q\>}\,
\\
\noalign{for $Q=A,\Omega,X$, while}
\Delta \Phi=\frac{\<e^{i\Theta}\_y_\Phi,\,\Du\>}{\<\_y_\Phi,y_\Phi\>}
  - X\,\frac{\<e^{i\Theta}\_y_\Omega,\,\Du\>}{\<\_y_\Omega,y_\Omega\>}\,.
\end{gather}
\end{subequations}
All of these results reduce to the standard perturbation
theory for the NLS equation 
(e.g., see Refs.~\cite{Iannone,MBK:SJAM})
when $s=0$.
The connection between invariances, conservation laws, and linear modes
of the DMNLS equation is further explored in Ref.~\cite{noether}.
In section~\ref{s:is} we will use the linear modes and Eq.~\eqref{e:intlchng} 
to guide the simulations of noise-perturbed lightwave systems.

\section{Rare events in dispersion-managed lightwave systems}
\label{s:is}

As mentioned in the introduction, the performance of 
both optical communication systems and certain femtosecond lasers
is limited by noise.
As a special but physically important example, we therefore
now use the results of section~\ref{s:dmnls} to quantify the 
effects of noise on dispersion-managed solitons.
Specifically,
we take $h(x,t)$ in Eqs.~\eqref{e:PNLS+DM} and~\eqref{e:PDMNLS} to be
\begin{equation}
h(x,t)= \sum_{n=1}^{N_a} \nu_n(x)\,\delta(t-nt_a)\,,
\end{equation}
where $t_a$ is the dispersion map period, as before,
and $\delta(t)$ is the Dirac delta distribution.
In other words, $u(x,t)$ evolves according to the unperturbed 
DMNLS equation~\eqref{e:DMNLSx} at all times except at $t= nt_a$,
for $n=1,\dots,N_a$, when
\begin{equation}
u(x,nt_a^+)=u(x,nt_a^-) + \epsilon\,\nu_n(x)\,,
\end{equation}
where $\nu_n(x)$ is taken to be normalized white Gaussian noise,
satisfying
\begin{equation}
\E[\nu_n(x)]=0\,,\qquad
\E[\nu_n(x)\nu^*_{n'}(x')]= \delta(x-x')\,\delta_{nn'},
\label{e:noisevar}
\end{equation}
where $\E[\,\cdot\,]$ denotes ensemble average,
and where in this case the small parameter $\epsilon^2$ 
is the dimensionless noise variance.
Starting from a soliton input pulse, namely $u(x,0)$ given by 
Eq.~\eqref{e:trvsln},
we are then interested in computing the probability density function 
(pdf) of the soliton parameters at the output time~$N_at_a$.

Even if the statistics of the noise sources are Gaussian, 
the resulting statistics at the output is not, in general, 
because propagation is nonlinear. 
Indeed, as mentioned earlier,
the combination of noise and nonlinearity presents 
a formidable challenge if one is interested in calculating the
probabilities of errors when performance standards dictate that
errors be rare events. 
It has recently been shown that 
variance reduction methods such as importance sampling 
can be used to calculate pdfs in such systems accurate to 
very small probabilities~\cite{JLT22p1201,OL28p105,OC256p439,PTL18p886}.
Here we use the results of section~\ref{s:dmnls} to implement 
IS methods for the DMNLS equation in the presence of noise.

\subsection{Importance sampling for the DMNLS equation}

The idea behind importance sampling is straightforward:
in order to calculate the probability of a desired rare event, 
sample the noise from a biased distribution 
that makes the rare events occur more frequently 
than would naturally be the case, while simultaneously 
correcting for the biasing.

Consider a set of random variables~$\~X=(x_1,\dots,x_N)$ 
distributed according to a joint probability distribution $p(\~X)$.
The probability $P$ that a function $y(\~X)$
falls into a desired range~$Y_d$
can be expressed via the multi-dimensional integral
\begin{subequations}
\label{e:prob}
\begin{equation}
P=\P[y \in  Y_d] =  \E[I(y(\~X))]= \int I(y(\~x)) p(\~x) d\~x\,,
\end{equation}
where the \textit{indicator function} $I(y)$ 
equals~1 when $y\in Y_d$ and 0~otherwise.
An unbiased estimator for~$P$ can be constructed via 
Monte-Carlo (MC) quadrature as
\begin{equation}
\^P= \frac1M \sum_{m=1}^M I(y(\~X_m)),
\label{e:MCprob}
\end{equation}
\end{subequations}
where the $M$ samples $\~X_m$ are drawn from the distribution $p(\~X)$. 
If $P$ is very small, however,
an unreasonable number of samples are necessary to produce
events for which $y$ is in~$Y_d$, 
let alone enough to accurately approximate the integral.  
However, one can rewrite Eqs.~\eqref{e:prob} 
as follows:
\begin{subequations}
\label{e:isprob}
\begin{gather}
\P[y \in  Y_d] =  
  \int I(y(\~x)) L(\~x)p_*(\~x) d\~x\,,
\\
\^P_*= \frac1M \sum_{m=1}^M I(y(\~X_{*,m}))L(\~X_{*,m})\,,
\end{gather}
\end{subequations}
where the samples $\~X_{*,m}$ are now drawn from the 
\textit{biasing distribution}~$p_*(\~X)$, and where  
the quantity $L(\~X)= p(\~X)/p_*(\~X)$ is the \textit{likelihood ratio}.
When an appropriate biasing distribution is selected,
importance-sampled Monte-Carlo (ISMC) simulations 
can accurately estimate the probability of the sought-after rare events
with a small fraction of the number of samples that
would be necessary with straightforward MC methods.
The challenge is, of course, to properly choose the biasing distribution.
Indeed, in order for importance sampling to work,
$p_*(\~X)$ should preferentially concentrate the MC samples 
around the \textit{most likely} system realizations that 
lead to the rare events of interest.
In our case the random variables~$\~X$ are the noise components 
added at the end of each dispersion map period.
Thus, in order to successfully apply IS we must find 
the most likely noise realizations that lead to 
a desired value of the soliton parameters at the output.

By substituting $\nu_n(x)$ into Eq.~\eqref{e:intlchng} 
of section~\ref{s:change}, we immediately obtain the 
noise-induced parameter change at the $n^\mathrm{th}$ map period as:
\begin{equation}
\Delta Q_n=
  \frac{\Re \int \e^{-i\Theta(x)}\_y^*_Q(x)\nu_n(x)\, dx}
    {\int\_y^*_Q(x) y_Q(x) dx}\,.
\label{e:Qchg}
\end{equation}
where $Q=A,\Omega,X$ as before, with a slightly more complicated
expression for $\Delta\Phi_n$.
Moreover,
for white Gaussian noise, maximizing the probability of a specific
noise realization
is equivalent to minimizing the negative of the argument of 
the exponential in the pdf; that is, minimizing the quantity
\begin{equation}
\int |\nu_n(x)|^2dx\,.
\label{e:minbiasnorm}
\end{equation}
Hence, in our case, the problem of determining the optimal biasing 
amounts to finding the noise realization
that minimizes the integral in Eq.~\eqref{e:minbiasnorm}
subject to the constraint of achieving a desired parameter change, 
that is, 
subject to the constraint $\Delta Q_n={}\Delta Q_{\mathrm{target}}$,
with $\Delta Q_n$ given by Eq.~\eqref{e:Qchg}.
This optimization problem can be solved by formulating
Eqs.~\eqref{e:minbiasnorm} and~\eqref{e:Qchg}
as a Lagrange multiplier problem, 
as in Refs.~\cite{OL28p105,PTL17p1022}. 
Solving the resulting problem then yields 
$\nu_n(x)=\nu_{n,\mathrm{opt}}(x)$,
where
\begin{eqnarray}
\nu_{n,\mathrm{opt}}(x)= 
  \frac{\Re \int \_y^*_Q(x') y_Q(x') dx'}{\int|\_y_Q(x')|^2dx'} \,
  \Delta Q_{\mathrm{target}}\,\e^{i\Theta(x)} \_y_Q(x)\,.
\label{e:optbias}
\end{eqnarray}
To induce a larger than normal parameter change,
we then concentrate the MC samples around this optimal path.
We do so by biasing the noise adding $\nu_{n,\mathrm{opt}}(x)$ 
as a deterministic component;
that is, we take
\begin{equation}
\nu_{*,n}(x)=\nu_{n,\mathrm{opt}}(x)+\upsilon_n(x)\,.
\label{e:biasednoise}
\end{equation}
where $\nu_{n,\mathrm{opt}}(x)$ is given by Eq.~\eqref{e:optbias},
and where $\upsilon_n(x)$ is also a white noise process satisfying
Eqs.~\eqref{e:noisevar}.

Note that the optimal path $\nu_{n,\mathrm{opt}}(x)$ depends on 
both the eigenmodes and the adjoint eigenmodes of the 
linearized DMNLS equation found in section~\ref{s:dmnls}.
In particular, Eq.~\eqref{e:optbias} implies that 
the optimal biasing is proportional to the \textit{adjoint eigenmode}
of the quantity that one desires to change,
a result that might not be obvious apriori.

Note also that, once the most likely noise realization that produces a 
given parameter change $\Delta Q_n$ at each map period are known,
one should also find the most likely way to distribute
a total parameter change $\Delta Q_{\mathrm{tot}}$ at the output 
among all map periods (cf. Refs.~\cite{MBK:SJAM,noether,PTL17p1022}).
This further optimization problem can also be formulated 
and solved as a variational problem~\cite{noether}.
When targeting large amplitude or frequency changes, however, 
it suffices to simply distribute 
equally this total change among all map periods.  
Thus, 
in the numerical simulations described in section~\ref{s:simulations}
we set $\Delta Q_{\mathrm{target}}= \Delta Q_{\mathrm{tot}}/N_a$ 
for $n=1,\dots,N_a$.

\subsection{Importance-sampled Monte-Carlo simulations}
\label{s:simulations}

We now apply the ideas presented above to concrete numerical
experiments of dispersion-managed systems under the effect of noise.
We perform importance-sampled Monte-Carlo (ISMC) simulations of 
the DMNLS equation~\eqref{e:DMNLS} perturbed by noise, 
and we compare the results to standard Monte-Carlo simulations of 
the original NLS equation with dispersion management~\eqref{e:NLS+DM},
also subject to noise.

Let us discuss the approach we used for the numerical simulations
of the noise-perturbed DMNLS equation.
We simulate the evolution of a dispersion-managed optical signal
by solving Eq.~\eqref{e:DMNLSf} numerically and adding noise to
the signal at periodically spaced time intervals.
The initial condition, i.e., the input DMS shape at~$t=0$,
is generated by solving the nonlinear integral equation
Eq.~\eqref{e:DMNLSintleqn} as explained in Appendix~\ref{s:A2},
and time evolution is performed with a fast numerical method,
as described in Appendix~\ref{s:A1}.  
White noise, which is added to the signal at $t=n\/t_a$ for $n=1,...,N_a$,
is numerically discretized as a collection of 
independent, identically distributed zero-mean normal random variables, 
one each for the real and imaginary parts of the signal at 
each spatial grid point.
Propagation and the addition of noise continue in this way
until the signal reaches the output at $t_{\mathrm{out}}=N_at_a$.

In standard Monte-Carlo simulations, one repeats the above
process for several different noise realizations
while monitoring the output value of the quantities of interest
(e.g., energy and/or frequency), and then computes their statistics.

For importance-sampled Monte-Carlo simulations, 
one also uses the basic framework described above.  
If one wants to obtain larger-than-normal deviations
of a quantity~$Q$, however,
one also performs the following steps at each map period 
before adding the noise:
\begin{enumerate}
    \item 
    Determine the underlying DMS from the noisy signal.  We do this by
    using the noisy pulse as the initial condition in the iteration
    scheme for the nonlinear integral equation~\eqref{e:DMSev}, 
    where the denominator in the RHS is replaced by
    $\lambda+\dbar(\omega-\Omega)^2$ to automatically obtain a DMS shape
    with the correct carrier frequency.  
    The values of $\Omega$ and $\lambda$ are obtained from the 
    (low-pass filtered) noisy signal,
    respectively by using moments and by matching the total energy in
    a pre-computed table of DMS energy as a function of~$\lambda$.
    \item 
    Obtain the linear modes and adjoint modes of the linearized DMNLS
    equation around the given DMS.
    We do this by numerically calculating the $x$ and $s$ derivatives 
    of the underlying DMS, and then using Eqs.~\eqref{e:Lambdamodes}
    and~\eqref{e:adjmodes}.
    The $x$-derivative is calculated using pseudo-spectral methods,
    while the derivative with respect to~$s$ is calculated by performing 
    step~1 twice, once at the given value of $s$ and once at~$s+ds$.
    \item  
    Generate an unbiased noise realization, 
    shift its mean with the appropriately scaled 
    adjoint mode associated with~$Q$
    according to Eqs.~\eqref{e:biasednoise} and~\eqref{e:optbias}, 
    and update the likelihood ratio.
\end{enumerate}
One then adds the noise to the pulse, propagates the noisy signal 
to the next map period, and repeats this process until the 
signal reaches the output.  
For a given simulation, several thousand ISMC samples,
generated with a few different biasing targets, are collected,
and their contributions are combined using multiple importance sampling 
\cite{JLT22p1201,Veach}
in order to numerically generate the pdf of the quantity of interest. 

\begin{figure}[t!]
\centerline{\includegraphics[width=0.45\textwidth]{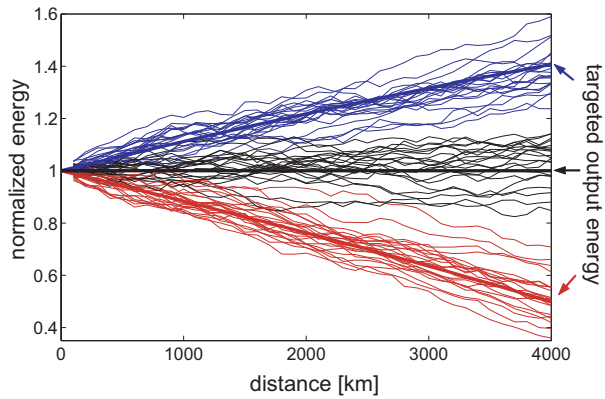}\kern-0.4em}
\caption{Samples from ISMC simulations of the DMNLS equation.  
Here, the pulse energy (normalized to input energy) 
is plotted as a function of time (i.e., distance in physical units).
The arrows represent the different targeted output energies:  
a larger than normal output energy (blue),
a smaller than normal output energy (red),
and unbiased energy (black). 
Also plotted are deterministic paths (thick, smooth curves, with 
color corresponding to the target) predicted by our perturbation theory. 
These are the preferential paths around which we attempt to sample 
by biasing the noise with the adjoint linear modes.
For each of three different targeted output energies,
a few dozen ISMC samples are also shown (also colored correspondingly),
demonstrating that the actual trajectories indeed follow the predictions 
of the theory.
See text for a detailed discussion of system parameters.}
\kern-\smallskipamount
\label{f:energysamples}
\kern-\medskipamount
\end{figure}

Even though at each map period the noise induces 
only a small change in the solution parameters, 
these small changes can accumulate into large parameter changes 
at the output, resulting in a significantly distorted signal.  
This gradual build up of noise-induced distortions 
is evident in Figure~\ref{f:energysamples}, where 
we plot the energy as a function of time for several different
noise realizations biased around the optimal energy path 
predicted by the theory.
For each sample path in this figure, the noise added at each map period
is biased by adding a proper multiple of the adjoint amplitude mode in
order to change the pulse amplitude and hence its energy.
Note how the random samples are concentrated near the trajectories
predicted by the theory.
One can think of these trajectories as a low-dimensional projection of 
a near-optimal path through state space to reach a targeted rare event.

Note also that the linearized DMNLS equation is only used to guide
the biasing, while, for each individual sample in the ISMC simulations, 
the full DMNLS equation is solved to propagate the signal.  
That is, no approximations are used for propagating the signal, 
and no assumptions are made about its shape to predict or calculate 
the pulse parameters at output.  
In other words, the only approximation in the simulations 
(beyond roundoff and truncation due to discretization) 
lies in using the information based on the linearized DMNLS equation 
in order to bias the noise.
Thus, use of importance sampling enables full nonlinear simulation 
of large, noise-induced parameter changes.

\subsection{Results and discussion}

As a test of the method, we performed numerical simulations of
two noise-perturbed dispersion-managed systems with different sets 
of physical parameters.
For both systems we compared ISMC simulations of the 
DMNLS equation~\eqref{e:DMNLS} 
to standard MC simulations of the original NLS equation with 
dispersion management~\eqref{e:NLS+DM},
numerically integrated with a standard second-order split-step
Fourier method.
For both systems we take a piecewise constant dispersion map.  
That is, we consider the transmission link to be comprised of 
alternating sections of fiber with opposite signs of dispersion. 

In the first system, which we will refer to as system~(a), 
we seek large changes in frequency at the output. 
In the second system, referred to hereafter as system~(b), 
we seek large changes in amplitude and hence in output energy. 
In both cases we choose the system parameters based on 
realistic values for optical fiber communication systems.
Typical values of system parameters for the DMNLS as a model of 
femtosecond lasers can be obtained from Ref.~\cite{Qurashi}.
We consider an average dispersion of 0.15\,ps$^2$/km
a nonlinear coefficient of 1.7\,(W$\cdot$km)$^{-1}$, 
and a fiber loss of 0.21\,dB/km.
We set a unit time of 17\,ps,
normalizing~$x$ in Eq.~\eqref{e:NLS+DM} correspondingly,
and we use the resulting dispersion length of 1,923\,km
to normalize~$t$,
resulting in $\dbar=1$ in Eq.~\eqref{e:NLS+DM}.
We further space amplifiers every 100\,km
(resulting in~$t_a=0.052$),  
taking the dispersion map period to be aligned with 
the amplification period.
The corresponding power needed to have $\=g=1$ in Eq.~\eqref{e:NLS+DM}
is 2.96\,mW, and we use this value as a unit to normalize pulse powers.
For system~(a) we consider a map strength of $s=2$ 
and a propagation distance of 2,000\,km (resulting in $N_a=20$).  
For system~(b) we consider a map strength of $s=4$ 
and a propagation distance of 4,000\,km (resulting in $N_a=40$).  
Thus, in both systems the average dispersion is small, while
the local dispersion is large in magnitude. 
[Recall that the map strength parameter~$s$ quantifies
the difference in magnitude between local and average 
values of dispersion, cf. Eq.~\eqref{e:sdef}.] 
For system~(a) we take $\lambda=1.5$
(corresponding to an initial pulse with 6.66\,mW peak power)
and for system~(b) we take $\lambda=2$ 
(corresponding to 11.8\,mW).
Finally, assuming a spontaneous emission factor of 1.5,
system~(a) has an optical signal-to-noise ratio of~16.7
(resulting in a dimensionless noise variance $\epsilon^2=2.372\cdot10^{-4}$),
system~(b) of~13.8 (resulting in $\epsilon^2=9.486\cdot10^{-4}$).

The output distributions of both frequency and energy are of course 
of practical interest in communications, since large deviations 
of each quantity will result in transmission errors.
Frequency changes translate in group velocity changes, 
and hence in the pulse walking off its assigned bit slot.
Similarly, a pulse that loses a significant fraction of 
its energy will be incorrectly detected in an amplitude-shift-keyed system.

\begin{figure}[t!]
\centerline{\includegraphics[width=0.4305\textwidth]{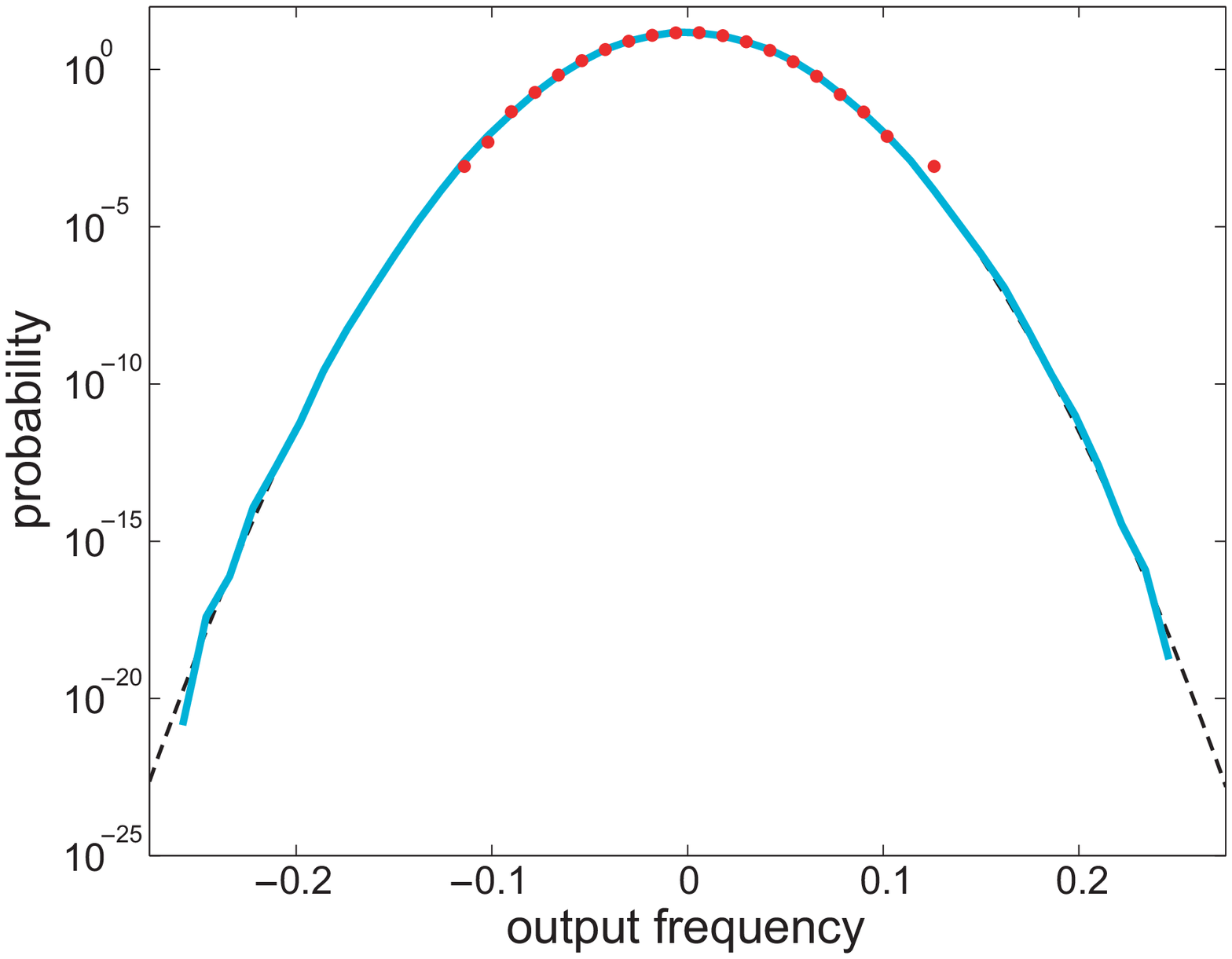}}
\caption{Probability density function of output frequency 
in a dispersion-managed system.
The solid cyan curve shows results from ISMC simulations 
of the DMNLS Eq.~\ref{e:DMNLS} with 75,000 samples.
The red dots result from standard MC simulations 
of the NLS equation with DM, Eq.~\eqref{e:NLS+DM} with 100,000 samples. 
The black curve is a Gaussian pdf obtained from Gordon-Haus theory 
of the NLS equation with DM.
Note how unbiased MC simulations of the NLS equation with DM 
agree exactly with the ISMC simulation of the DMNLS and the 
Gaussian fit to that simulation as far down in probability
as the unbiased simulations can reach.}
\label{f:freqpdf}
\vskip1.5\bigskipamount
\centerline{\includegraphics[width=0.4305\textwidth]{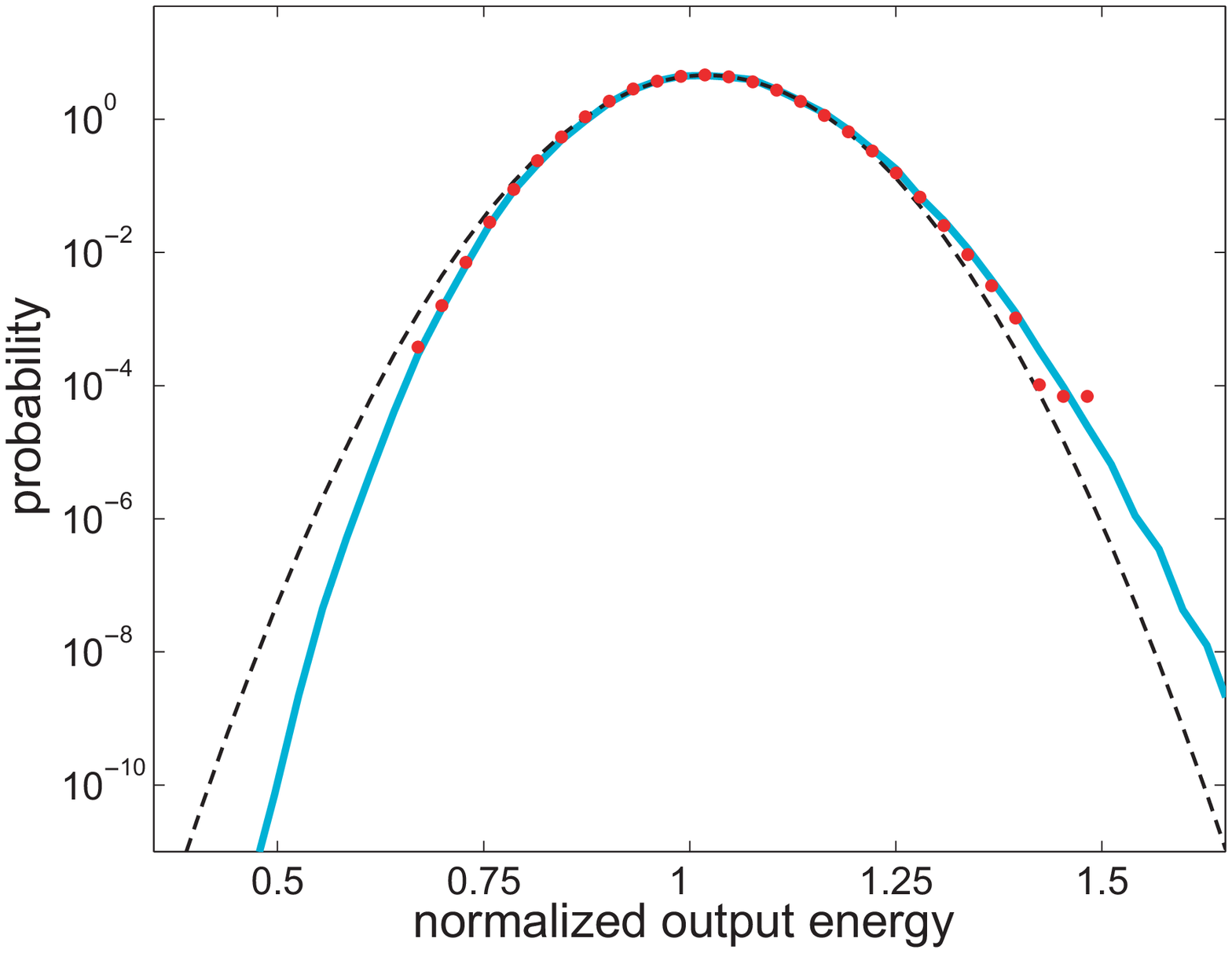}}
\caption{Probability density function of normalized output energy. 
The solid cyan curve shows results from ISMC simulations of the DMNLS
with 42,000 samples. 
The red dots result from standard MC simulations 
of the NLS equation with DM with 1,000,000 samples. 
The black curve is a Gaussian fit to that simulation.  
Note how unbiased MC simulations of the NLS equation with DM 
clearly deviate from Gaussian, but agree well with ISMC simulations 
of the DMNLS as far as down in probability as the unbiased simulations 
can accurately reach.}
\label{f:enrgpdf}
\kern-\medskipamount
\end{figure}

In Figure~\ref{f:freqpdf} we plot the pdf of the output frequency 
of pulses from the dispersion-managed system~(a) described above. 
We perform standard MC simulations of the noise-perturbed 
NLS equation with DM~\eqref{e:PNLS+DM}, with 100,000 samples 
to estimate the pdf of output frequency (red dots). 
We also plot a Gaussian pdf (black dashed line) whose variance is 
consistent with a theoretical model 
of Gordon-Haus effect for a DM system governed by the noise-perturbed 
NLS equation in the presence of dispersion management~\eqref{e:PNLS+DM}
\unskip~\cite{OL22p1870,OC200p165,EL32p2085},
assuming a Gaussian ansatz for the pulse shape at the chirp-free point.
Finally, we perform ISMC simulations of the noise-perturbed DMNLS 
equation~\eqref{e:PDMNLS} with 75,000 samples, and we plot the
the corresponding results for the pdf of output frequency (cyan solid line).  
The results of ISMC simulations of the DMNLS equation, 
of standard MC simulations of the NLS equation with DM, 
and the Gaussian fit to that simulation all match exactly 
to very small probabilities.  
This comparison 
demonstrates the effectiveness of 
using the modes of the linearized DMNLS equation to find rare events 
in dispersion-managed optical systems.

In Figure~\ref{f:enrgpdf} we plot the pdf of the output energy 
of pulses in dispersion-managed system~(b) described above. 
Here, as in Fig.~\ref{f:energysamples},
energy is normalized by the the energy of the input signal,
i.e., by the ``back-to-back" signal energy.  
Again, the red dots show results of standard MC numerical simulations 
of the noise-perturbed NLS equation with DM, with 1,000,000 samples;
the black dashed line shows a Gaussian fit to the results of these 
simulations,
and the cyan solid line shows the results of ISMC simulations 
of the noise-perturbed DMNLS equation, with 42,000 samples.
It is worth noting that the pdf generated from standard MC simulations 
of the NLS equation with DM clearly deviates from the Gaussian fit,
but it agrees very well with the ISMC simulations of the DMNLS equation,
as far down in probability as the unbiased MC simulations can reach.
These comparison provides a strong validation of the DMNLS equation
as a model of dispersion-managed lightwave systems.

\section{Conclusions}

We described a perturbation theory for soliton-based
dis\-persion-managed lightwave systems, whose dynamics 
is governed by the dispersion-managed NLS (DMNLS) equation, 
and we used the results of the theory to guide 
importance-sampled Monte-Carlo simulations to quantify the 
effects of noise in these systems.  
The present theory differs from the soliton perturbation theory 
which applies to the NLS equation in several important respects.
First, due to the loss of integrability, the eigenmodes of the 
linearized DMNLS equation are derived from the invariances of 
the equation rather than from the inverse scattering method.
Second, the DMNLS equation is not scale-invariant, 
but is invariant under a generalized scaling transformation;
as a consequence the amplitude mode depends explicitly 
on the map strength.
Third, unlike for the NLS equation,
the linear modes of the DMNLS equation are not automatically bi-orthonormal, 
and in particular their norms and inner products depend on 
both the map strength and the pulse energy.

The results of importance-sampled numerical simulations of the 
noise-perturbed DMNLS equations agree very well with the results
of Gordon-Haus theory for dispersion-managed systems (which are
based on the original NLS equation) as well with the results 
of standard Monte-Carlo simulations of the original NLS equation 
with dispersion management as far down as those can go in probability.
This is true even when those results deviate from Gaussian distributions.
Both of these results provide a further important test of 
the validity of the DMNLS equation as a model of 
dispersion-managed lightwave systems.

It should be noted that, in some parameter regimes, 
the DMNLS equation also admits some internal modes~\cite{Capobianco,PRE62p4283}.
The accumulation of noise components in such internal modes 
could also contribute to system failures.
If no generalized modes are associated with these modes,
however, the variance of the resulting noise-induced pulse fluctuations 
would grow linearly in time, as opposed to cubically
(as is the case for phase and timing fluctuations).
Therefore, one would expect these modes not to be the dominant
source of errors.
Of course, this argument should be validated by more precise calculations
and/or careful numerical experiments.  
We will address this issue in the future.

\section*{Acknowledgments}

It is a pleasure to thank M.~J.~Ablowitz, W.~L.~Kath and 
C.~R. Menyuk for many interesting discussions.
This work was supported by the National Science Foundation under grant 
DMS-0506101.

\appendix
\section*{Appendix: Fast numerical methods for the DMNLS equation}
\setcounter{section}1
\renewcommand\theequation{\Alph{section}.\arabic{equation}}

\noindent
Here we discuss fast numerical methods for the DMNLS equation~\eqref{e:DMNLS} 
which were used in our numerical simulations. 
We address two issues:
(i)~numerical methods for time integration,
and (ii)~numerical methods to find traveling-wave solutions.

\subsection{Time integration of the DMNLS equation}
\label{s:A1}

We first discuss time integration methods for the DMNLS equation.
Equation~\eqref{e:DMNLSx} differs from a standard PDE because of 
the double integral, and it should be obvious that 
the most computationally expensive task when trying to integrate it numerically
is the evaluation of the double integral in either of Eqs.~\eqref{e:DMNLS}.
If $N_x$ is the number of grid points in $x$ (or $\omega$),
a straightforward quadrature scheme requires $O(N_x^3)$ operations
(since one needs to evaluate a different double integral 
for each value of $x$ or $\omega$).
As we now show, however,
it is possible to evaluate the integral with only $O(JN_x\log N_x)$
operations, 
where $J$ is in principle an arbitrary constant
whose meaning will become clear shortly.

Let us denote the double integral in Eq.~\eqref{e:DMNLSf} as
\begin{equation}
\^K(\omega,t)=
  \iint   
  \^u'_{(\omega+\omega')}\^u'_{(\omega+\omega'')}\^u'^*_{(\omega+\omega'+\omega'')}
  r_{(\omega'\omega'')}\, \d\omega'\d\omega''\,
\end{equation}
(where we reinstated the primes to distinguish the solution of
the DMNLS equation from that of the original NLS problem).
Recall from Eq.~\eqref{e:solndecomposition} that 
$\^u(\omega,t,\zeta)= \^u'(\omega,t)\,\e^{-iC(\zeta)\omega^2/2}$ 
is the leading-order solution 
of the original NLS equation with dispersion management, 
namely Eq.~\eqref{e:NLS+DM}.
Also recall from~Eq.~\eqref{e:rdef} that the kernel $r(x)$ is an average 
over a dispersion map period,
and $R(t,t')= \F^{-1}_{t,t'}[r(\omega\omega')]$.
Indeed, retracing backwards the steps of the 
multiple scale expansion used to obtained the DMNLS equation~\cite{OL23p1668}, 
one can express the whole quantity $\^K(\omega,z)$ as an average:
\begin{equation}
\^K(\omega,t)=
  \int_0^1 \e^{iC(\zeta)\omega^2/2}\,g(\zeta)
    \,\F_\omega[(|u|^2u)(x,t,\zeta)] \,\,d\zeta\,,
\label{e:Ndef}
\end{equation}
where $u(t,z,\zeta)=\F^{-1}_t[\^u(\omega,z,\zeta)]$.
Note that~$t$ is a constant in the integral on the RHS of Eq.~\eqref{e:Ndef},
and this is precisely the key for the fast numerical computation 
of~$\^K(\omega,t)$.

Divide the interval $[0,1]$ into $J$ equally spaced points
$\zeta_0,\dots,\zeta_J$, with $\zeta_0=0$ and $\zeta_J=1$.
For each fixed value of~$t$
we can calculate $\^K(\omega,t)$ as follows:
\begin{enumerate}
\itemsep0pt\parsep0pt
\item
Fix $\zeta_j$ and construct 
$\^u(\omega,t,\zeta_j)=\^u'(\omega,t)\,\e^{-iC(\zeta_j)\omega^2/2}$
\item
Take the inverse Fourier transform to obtain $u(x,t,\zeta_j)$ 
and construct the product $|u|^2u$. 
\item
Take the direct Fourier transform to obtain $\F_\omega[|u|^2u]$.
\item
Multiply the result by $g(\zeta_j)\,\e^{iC(\zeta_j)\omega^2/2}$ to obtain the 
integrand in Eq.~\eqref{e:Ndef}.
\item
Repeat steps 1--4 for all grid points $\zeta_j$ and 
evaluate the integral in Eq.~\eqref{e:Ndef} to obtain $\^K(\omega,t)$.
\end{enumerate}
The DMNLS equation can now be integrated in time 
using any desired numerical scheme.
For example, 
one can use an exact integrating factor on the linear part 
of Eq.~\eqref{e:DMNLS}
and an explicit fourth-order RK4 for the nonlinear part.
%
A few remarks are now in order:
\leftmargini 1.25em
\begin{itemize}
\itemsep 0pt
\item[\circ]
The above procedure,
which can be carried out for any choice of dispersion map 
and nonlinear coefficient, is a generalization of an algorithm 
originally introduced in Ref.~\cite{OL26p1535}.
\item[\circ]
Steps 2 and 3 each cost $O(N_x\log N_x)$ operations.  
Since they must be repeated for each of the grid points $\zeta_0,\dots,\zeta_J$, 
the overall complexity of the algorithm is $O(JN_x\log N_x)$.
\item[\circ]
In essence, the algorithm reduces the calculation of $\^K(\omega,z)$ to 
that of the effective nonlinearity in the original NLS equation, 
averaged over one period of the dispersion map.  
\item[\circ]
In practice, the value of $J$ is dictated by the value of the map strength
and the need to adequately resolve the changes in the integrated dispersion 
function~$C(\zeta)$.
The same requirements however also dictates the integration step size 
in the original NLS problem.
The computational complexity of the DMNLS equation~\eqref{e:DMNLS} 
is thus the same as that of the original NLS problem~\eqref{e:NLS+DM}.  
\item[\circ]
It appears more advantageous to integrate Eq.~\eqref{e:DMNLSf} 
rather than Eq.~\eqref{e:DMNLSx}, since this allows
one to treat the linear (stiff) portion of the PDE exactly.
Also, 
the use of a split-step method does not seem as desirable here
(unlike the case of the NLS equation),
since it is not possible to integrate the nonlinear portion exactly.
\end{itemize}

\subsection{Traveling-wave solutions of the DMNLS equation}
\label{s:A2}

We now discuss a numerical method to find stationary solutions of the 
DMNLS equation, i.e., dispersion-managed solitons (DMS).
That is, we look for solutions of 
Eq.~\eqref{e:DMNLSx} in the form
\begin{subequations}
\begin{equation}
u'_\mathrm{st}(x,t)= f(x)\,\e^{i\lambda^2t/2}\,,
\end{equation}
or, equivalently, solutions of Eq.~\eqref{e:DMNLSf} in the form
\begin{equation}
\^u'_\mathrm{st}(\omega,t)= \^f(\omega)\,\e^{i\lambda^2t/2}\,.
\label{e:DMSo}
\end{equation}
\end{subequations}
Once such a stationary solution is available, 
a three-parameter family of travelling waves can be generated 
using the translation, phase and Galilean invariance of 
the DMNLS equation.
That is, if $u'_\mathrm{st}(x,t)$ 
is any solution of Eq.~\eqref{e:DMNLSx}, so is
\begin{equation}
u'_{\Omega,x_o,\phi_o}(x,t)= 
  \e^{i[\Omega x - \dbar\Omega^2 t/2 \, + \phi_o]}\,
  u'_\mathrm{st}(x-x_o-\dbar\Omega\,t,t)\,,
\label{e:DMSfamily}
\end{equation}
where $\Omega$, $x_o$ and $\phi$ are arbitrary real parameters.

Inserting Eq.~\eqref{e:DMSo} into the DMNLS equation
yields the nonlinear integral equation~\eqref{e:DMNLSintleqn},
which we rewrite here for convenience:
\begin{multline}
\^f_{(\omega)}= 
\\
  \frac2{\lambda^2+\dbar\omega^2} \,\,
  \iint   
  \^f_{(\omega+\omega')}\^f_{(\omega+\omega'')}\^f^*_{(\omega+\omega'+\omega'')}
  r_{(\omega'\omega'')}\,\d\omega'\d\omega''\,. 
\label{e:DMSev}
\end{multline}
For each fixed value of $\lambda$, 
Eq.~\eqref{e:DMSev} yields the shape
of the corresponding dispersion-managed soliton.
(Or, rather, its Fourier transform.)\,\ 
Thus, for each value of $s$, 
$\lambda$ plays the role of a nonlinear eigenvalue,
which is in one-to-one correspondence with the 
dispersion-managed soliton's energy.
For the NLS equation, one simply has 
$u(x,t)= A \sech[A x]\e^{iA^2t/2}$;
thus $\lambda=A$ is exactly half of the pulse energy:
$\int |u_s(x,t)|^2\,\d x=2A$\,.
This is related to the existence of a simple scaling invariance:
if $u_o(x,t)$ is any solution of NLS, so is
$u(x,t)= \lambda\,u_o(\lambda x,\lambda^2t)$.
When the map strength is non-zero, however, this invariance is lost,
and the scaling invariance of the DMNLS equation
is more complicated than that of the NLS equation,
as discussed in section~\ref{s:dmnls}.
As a consequence, 
a different integral equation~\eqref{e:DMSev} must be solved 
to obtain the soliton shape for each given value of~$\lambda$,
unlike the NLS equation.

Let $R[\^f_{(\omega)}]$ denote the right-hand side of Eq.~\eqref{e:DMSev}.
A naive approach to solving Eq.~\eqref{e:DMSev} is to simply apply 
a Neumann iteration scheme: $\^f_{(\omega)}\o{n+1}= R[\^f_{(\omega)}\o{n}]$.
This iteration scheme is divergent, however.
The key is to apply a modified iteration scheme:
\begin{equation}
\^f_{(\omega)}\o{n+1}= \big|C[\^f_{(\omega)}\o{n}]\big|^\alpha \,R[\^f_{(\omega)}\o{n}]\,,
\label{e:Neumann}
\end{equation}
where the convergence factor $C$ is used:
\begin{subequations}
\begin{gather}
C[\^f_{(\omega)}]= \frac{s_L[\^f_{(\omega)}]}{s_R[\^f_{(\omega)}]},
\label{e:convergence}
\\
\noalign{with}
s_L[\^f_{(\omega)}]= \int |\^f_{(\omega)}|^2\d\omega\,,\qquad
s_R[\^f_{(\omega)}]= \int \^f_{(\omega)}^* R[\^f_{(\omega)}]\,\d\omega\,.
\end{gather}
\end{subequations}
Again, a few remarks are in order:
\leftmargini 1.25em
\begin{itemize}
\itemsep 0pt
\item[\circ]
The method discussed in section~\ref{s:A1} for the fast 
numerical computation of the double integral in the DMNLS equation 
also applies, of course, for the calculation of the double integral 
in Eq.~\eqref{e:DMSev}.
\item[\circ]
Since $C[\^f_{(\omega)}]=1$ for all solutions $\^f_{(\omega)}$ 
of Eq.~\eqref{e:DMSev},
any solution of Eq.~\eqref{e:DMSev} is also a solution of 
the modified integral equation 
$\^f_{(\omega)}= |C[\^f_{(\omega)}]|^\alpha R[\^f_{(\omega)}]$,
for which Eq.~\eqref{e:Neumann} is a standard Neumann iteration scheme.
\item[\circ]
Since $C[\^f_{(\omega)}]=1$ when $\^f_{(\omega)}$ solves Eq.~\eqref{e:DMSev},
the value of $C[\^f_{(\omega)}]$ can be used as a monitor of convergence,
for example by requiring that 
$|1-C[\^f_{(\omega)}]|$ drop below a pre-defined threshold 
(e.g., $10^{-12}$ or $10^{-16}$)
as a termination condition.
\item[\circ]
This iteration scheme, which was first used in Ref.~\cite{OL23p1668} 
to find the shape of dispersion-managed solitons,
is based on a method introduced in Ref.~\cite{Petviashvili}.
A proof of the convergence of this method for a general class 
of nonlinear evolution equations was recently given in Ref.~\cite{Stepanyants}.
\item[\circ]
As might be expected, the choice of $\alpha$ is crucial.
Indeed, it can be shown that the method converges for $1<\alpha<2$, 
and optimal convergence is obtained for $\alpha=\frac32$.
Note, however, that\, $\dbar\ge0$ is required for convergence.
If $\dbar<0$ instead, the denominator of the RHS of Eq.~\eqref{e:DMSev}
has two simple poles at $\omega = \pm\lambda/\sqrt{|\dbar|}$, and 
even the modified iteration scheme diverges.
\end{itemize}

\catcode`\@ 11
\def\journal#1&#2,#3(#4){\begingroup{\sl #1\unskip}~{\bf\ignorespaces #2}\rm, #3 (#4)\endgroup}
\def\title#1{\textit{#1}}
\def\@biblabel#1{#1.}

\end{document}